\documentstyle[aps,pre,tighten,epsf]{revtex}

\begin{document}

\author{R. Reigada}
\address{Departament de Qu\'{\i}mica F\'{\i}sica,
Universitat de Barcelona, Avda. Diagonal 647, 08028 Barcelona, Spain}

\author{A. Sarmiento}
\address{ Instituto de Matem\'aticas, Universidad Nacional 
Aut\'onoma de M\'exico, Ave. Universidad s/n, 62200 Chamilpa,
Morelos, M\'exico}

\author{Katja Lindenberg} 
\address{Department of Chemistry and Biochemistry and Institute for
Nonlinear Science, University of California,
San Diego, La Jolla, California 92093} 

\title{Energy Relaxation in Nonlinear One-Dimensional Lattices}

\date{September 6, 2001}

\maketitle

\begin{abstract}

We study energy relaxation in thermalized 
one-dimensional nonlinear arrays of the Fermi-Pasta-Ulam type.
The ends of the thermalized systems are placed in contact with
a zero-temperature reservoir via damping forces.  Harmonic arrays relax
by sequential phonon decay into the cold reservoir, the lower
frequency modes relaxing first. 
The relaxation pathway for purely anharmonic arrays
involves the degradation of
higher-energy nonlinear modes into lower energy ones.  The lowest energy
modes are absorbed by the cold reservoir, but a small amount of energy is
persistently left behind in the array in the form of almost stationary
low-frequency localized modes.  Arrays with interactions that contain
both a harmonic and an anharmonic contribution exhibit
behavior that involves the interplay of phonon modes and breather
modes.  At long times relaxation is extremely slow due to
the spontaneous appearance and persistence of energetic high-frequency
stationary breathers.
Breather behavior is further ascertained by explicitly injecting a
localized excitation into the thermalized arrays and observing the
relaxation behavior.

\end{abstract} 

\bigskip

\noindent
PACS number(s): 05.40.-a, 05.45.-a, 63.20.Pw
\bigskip

%\begin{multicols}{2}
%\narrowtext

\section{Introduction}
\label{introduction}
The localization of vibrational energy in discrete nonlinear arrays
has attracted a huge amount of interest in the past 15 years as
a possible mechanism for the efficient storage and transport of
energy. 
The reasons for the broad interest in these phenomena are at least
two-fold: on the one hand, they embody many of the interesting effects
of the interplay of nonlinearity, discretization, and stochasticity, and
on the other they may be of use in explaining a variety of
physical and biophysical phenomena.  The interest in nonlinear arrays
serving as energy storage and transfer assemblies for chemical or
photochemical processes is not uncommon~\cite{various1},
and literature on the
subject goes back for two decades~\cite{various2}.  More recently, 
the localization and transport of vibrational
energy has been invoked in a number of physical settings including
DNA molecules~\cite{peyrard}, hydrocarbon structures~\cite{kopidakis}, the
creation of vibrational intrinsic localized modes in anharmonic
crystals~\cite{rossler}, photonic crystal waveguides~\cite{mingaleev}, and
targeted energy transfer between donors and acceptors in
biomolecules~\cite{aubry}.

Many types of nonlinear arrays exhibit spontaneous localization, and
in each of these the conditions that lead to localization are complex and
multifaceted.  A vast literature deals not only with different types of
arrays but with issues such as boundary conditions, initial conditions,
whether the system is closed or forced, thermal effects, etc.  It
is impossible to present here a full catalog or even sensibly organized
panorama of results; a number of excellent reviews
have aided greatly in the effort~\cite{FW,bri9}.

Increasing attention has recently been devoted to thermalized nonlinear
arrays. 
Whether energy can be transmitted without the destructive thermal and
dispersive degradation that occurs in harmonic lattices is clearly an
important question: one thinks, for example, about 
the amazing energy transfer cascade occurring in the photosynthetic
process~\cite{pho}, or about the ability of some proteins to efficiently
store and transport and thereby convert chemical energy into mechanical
energy~\cite{pro1,pro2,pro3}.  Transport across a thermal gradient
and questions concerning the validity of the usual Fourier law have
very recently been addressed in a number of
papers~\cite{hu,prosen,aoki,isbister}.  Relaxation
to thermal equilibrium and the nature of this equilibrium have been
addressed in recent
work~\cite{bourbonnais,piazza,harvesting,cretegny}, as has the
transport of energy in thermal
arrays~\cite{wepulse,wethermalresonance,burlakov,kosevich,floria}.

In addition to the phonons associated with the linear portion of the
potential in a nonlinear array, a variety of stationary, and non-stationary
but long-lived, excitations are possible, including
solitons~\cite{isbister,burlakov,kosevich,kruskal,dusi,kosevich2}
(long-wavelength excitations that persist from the continuum limit upon
discretization), periodic
breathers~\cite{FW,bri9,cretegny,burlakov,kosevich,dusi,kosevich2,bri2,page,sandusky,bri3,bri4,bri5,bri7,bri8}
(spatially localized time periodic excitations that
persist from the anticontinuous limit upon coupling), and so-called chaotic
breathers~\cite{cretegny} (localized excitations that evolve chaotically).  
These nonlinear excitations arise (even spontaneously)
and survive for a long time in numerical experiments, and they
clearly play an important role in determining the global
macroscopic properties of nonlinear extended systems. 

The nonlinearity in discrete nonlinear arrays may occur in the interactions
$V(x_i, x_j)$ and/or in the ``local" or ``external" potentials $U(x_i)$.
Here $x_i$ is the displacement of particle $i$ from its
equilibrium position.  While the presence of local potentials favors energy
localization, we are interested in localized energy that can also move,
and mobility tends to be easier in the absence of a local
potential~\cite{harvesting,wepulse,wethermalresonance}.  We
therefore focus on Fermi-Pasta-Ulam (FPU) lattices of
unit masses, each connected to its nearest neighbors by quadratic 
and/or quartic springs (FPU $\beta$-model).  Here we deal only with the
one-dimensional problem (the 
two-dimensional FPU system will be considered elsewhere~\cite{prep}).
The Hamiltonian of the system is
\begin{equation}
H = \sum_{i=1}^{N} \frac{\dot{x}_{i}^2}{2} + \frac{k}{2} \sum_{i=1}^{N}
(x_{i} - x_{i-1})^2 +\frac{k'}{4} \sum_{i=1}^{N} (x_{i} - x_{i-1})^4
\label{ham1}
\end{equation}
where $N$ is the number of sites.  $k$ and $k'$ are the harmonic
and anharmonic force constants, respectively.
In this work we focus on the thermal relaxation of FPU chains.
Thermal relaxation and associated mobility properties turn out to be
{\em entirely different} in purely harmonic ($k'=0$), purely anharmonic
($k=0$), and ``mixed" ($k$ and $k'$ $\neq 0$) $1D$ systems.

To study energy relaxation we initially thermalize the system
at temperature $T$ (see below).  We then connect the end sites of the system 
to a zero-temperature reservoir via appropriate damping terms and observe the
thermal relaxation of the array toward zero
temperature~\cite{piazza,tsi1,tsi2}.  In order to understand the role of
the various interactions (quadratic, quartic) and of the localized modes
that spontaneously emerge in the thermalization and relaxation process, we
perform a second numerical experiment
where we inject at the center of the thermalized chain a localized
breather-like excitation of energy much higher than the thermal energy.
The dynamics of such excitations have been studied in some detail in a
variety of contexts, but not in thermalized arrays.
Again, we observe how the thermal energy as well
as the excitation energy relax toward equilibrium. 

Section~\ref{initial} describes the preparation of our system.
In Sec.~\ref{thermal} we discuss the relaxation behavior of an initially
thermalized chain connected to a zero-temperature reservoir.
In Sec.~\ref{injected} we consider the relaxation behavior when a
high-energy localized excitation is introduced in the thermalized chain.
Section~\ref{summary} contains a summary of the results.  

\section{Initial Conditions}
\label{initial}
Different energy distributions in nonlinear arrays evolve quite
differently~\cite{bourbonnais,cretegny,dusi,deluca}, and
therefore existing work is not sufficient to predict the relaxation
behavior of initially thermalized arrays.
To thermalize the system to a given temperature $T$ we augment the equations
of motion resulting from Eqs.~(\ref{ham1}) with the Langevin
prescription connecting each site to a heat bath:
\begin{equation}
\ddot{x_i} = -\frac{\partial}{\partial x_i} [V(x_i -x_{i-1}) +
V(x_{i+1}-x_i)] -\gamma_0 \dot{x}_i +\eta_i(t).
\label{langfinitet}
\end{equation}
Here $V(x_i-x_j)$ is the FPU potential, $\gamma_0$ is the
dissipation parameter, and the $\eta_i(t)$ are mutually uncorrelated
zero-centered Gaussian
$\delta$-correlated fluctuations that satisfy the fluctuation-dissipation
relation at temperature $T$:
\begin{equation}
\langle \eta_i(t) \rangle = 0, \;\;\;\;\;\;\;\;\;\;\;\;
\langle \eta_i(t) \eta_j(t') \rangle = 2 \gamma_0 k_B T
\delta_{ij}\delta(t-t').
\label{fdr}
\end{equation}
The brackets here and below denote an ensemble average, and
$k_B$ is Boltzmann's constant.
We implement free-end boundary conditions, $x_0=x_1$ and $x_N=x_{N+1}$,
a common set of boundary conditions in relaxation
studies (we do stress, however, that although boundary conditions do
not strongly affect equilibrium properties, they do strongly affect some
relaxation dynamics).
For the integrations here and subsequently we use the fourth order
Runge-Kutta method.

The equilibrium energy landscape of our arrays can be characterized via
appropriate correlation functions and/or associated frequency spectra.
A convenient choice is the relative displacement autocorrelation function
\begin{equation}
C(\tau)= \frac{1}{(N-1)}\sum_{i=2}^{N} \left<
\Delta_i(t+\tau)\Delta_i(t)\right>
\label{equilC}
\end{equation}
where $\Delta_i(t)$ is the relative displacement
\begin{equation}
\Delta_i(t)\equiv x_i(t)-x_{i-1}(t).
\label{relative}
\end{equation}
The associated spectrum is
\begin{equation}
S(\omega)=2\int_0^{\infty} d\tau~ C(\tau) \cos \omega \tau.
\label{equilS}
\end{equation}

%figure 1
\begin{figure}[htb]
\begin{center}
\epsfxsize = 4.in
\epsffile{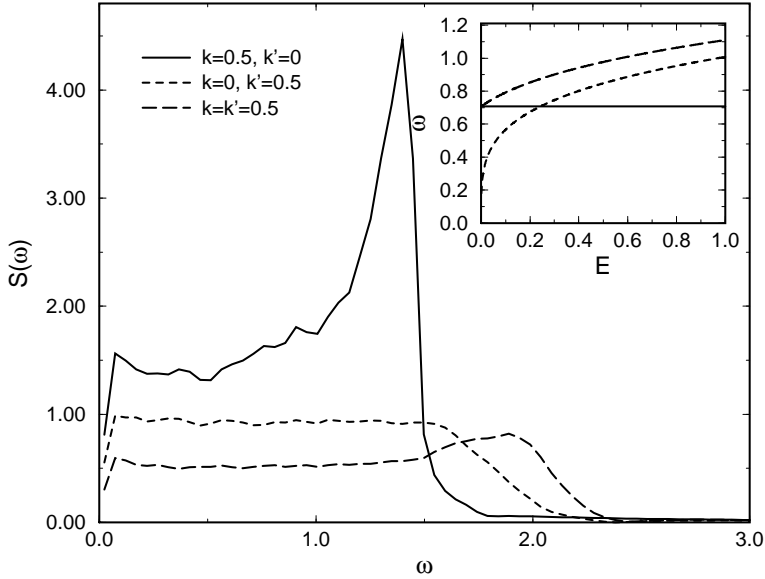}
\end{center}
\caption{Equilibrium spectrum for each of the three arrays with $N=50$
at $T=0.5$.  Inset: frequency vs. energy for the three potentials.}  
\label{figswt0}
\end{figure}

Typical equilibrium spectra for the three chains are shown in
Fig.~\ref{figswt0}.  
The temperature and other parameters are indicated in the figure and
caption.
The harmonic spectrum has a peak near $\omega=\sqrt{4k}=\sqrt{2}$, 
and this is also roughly the temperature-independent spectral width. 
This spectrum can be calculated analytically~\cite{wethermalresonance},
and one obtains (with periodic boundary conditions, but for sufficiently
long chains the boundary conditions don't affect the thermal equilibrium
spectrum):
\begin{equation}
S(\omega)=\frac{4\gamma_0 k_BT}{N} \sum_{q=0}^{N-1} \frac{1-\cos(2\pi
q/N)}{[r_1^2(q) +\omega^2][r_2^2(q)+\omega^2]}
\label{harmonicspectrum}
\end{equation}
where
\begin{equation}
r_{1,2}(q) = -\frac{\gamma_0}{2} \pm
\sqrt{\left(\frac{\gamma_0}{2}\right)^2 - 4k\sin^2\left(\frac{\pi
q}{N}\right)}.
\end{equation}

The inset in Fig.~\ref{figswt0} shows the frequency vs.
energy curves obtained from the
usual relation for the period of an oscillation in a potential $V(y)$,
\begin{equation}
\omega(E) = \frac{2\pi}{\tau(E)}, \qquad \tau(E) = 4\int_0^{y_{max}}
\frac{dy}{\sqrt{2[E-V(y)]}} ,
\end{equation}
where the amplitude $y_{max}$ is the positive solution of the equation
$V(y)=E$.  
Note that the frequency $\omega=\sqrt{k}$ associated with the harmonic
oscillator in the inset lies in the middle of this spectrum, and
that the harmonic
spectrum is temperature-independent except for the overall coefficient.
The temperature dependence of the nonlinear array spectra is considerably
more complex.
The purely hard FPU chain (short dashed lines) shows a broader spectrum,
consistent with the fact that at energies around 0.5 (viz. our
temperature), oscillator frequencies associated with
a purely hard potential are higher than that of a harmonic oscillator (cf.
inset).  For a considerably lower energy, say $E=0.1$, the typical frequency
associated with a purely hard oscillator in the
inset is {\em lower} than that of a harmonic oscillator; the
associated spectrum at a temperature $T=0.1$ (not shown here) is
narrower than that of the harmonic chain.
The mixed chain (long dashed lines) has a broader spectrum than the
harmonic or purely hard arrays at any temperature, again consistent with
the inset.  Whereas the harmonic
array only supports extended modes (phonons), some of the frequencies for 
the purely hard and mixed arrays 
are associated with nonlinear modes that include (especially at
high frequencies) localized modes.  Furthermore, it should be remembered
that whereas each phonon mode is characterized by a single frequency,
each nonlinear mode in general involves many frequencies. 

\section{Relaxation of Thermalized Arrays}
\label{thermal}

Relaxation of thermalized arrays from an initial temperature $T$
to zero temperature has been studied
in systems with nonlinear {\em local} potentials~\cite{tsi1,tsi2}. 
Some aspects of thermal relaxation in FPU chains have recently been
investigated~\cite{piazza}; our results significantly clarify and
expand on these recent results. In particular, we provide more detailed
insight into the mechanisms that contribute to the thermal relaxation
process.  We also provide a more detailed analysis of the relaxation
dynamics at long times.

%figure 2
\begin{figure}[htb]
\begin{center}
\epsfxsize = 4.in
\epsffile{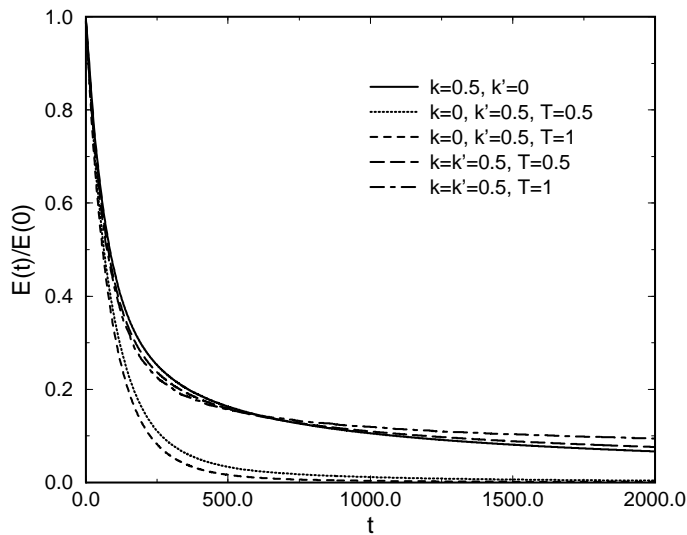}
\end{center}
\caption{Energy vs. time for various relaxing arrays with $N=50$.  Initially
each array is in thermal equilibrium at the temperature indicated in the
figure. The normalized energy of the harmonic array is independent of
temperature.}
\label{figdecay}
\end{figure}

\begin{figure}[htb]
\begin{center}
\epsfxsize = 4.in
\epsffile{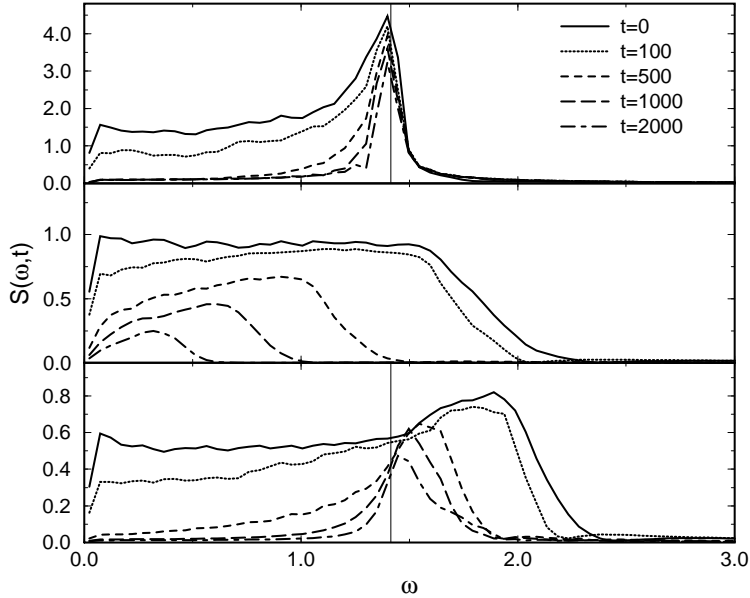}
\end{center}
\caption{Time evolution of spectra for various relaxing
arrays.  First panel: harmonic chain with $k=0.5$.  Second panel: purely
anharmonic chain with $k'=0.5$.  Third panel: mixed chain with $k=k'=0.5$.
The thin vertical line indicates the frequency
$\omega=\sqrt{4k} = \sqrt{2}$.
Initially ($t=0$ curves) each array is in thermal
equilibrium at $T=0.5$.}
\label{figsw}
\end{figure}

As in previous relaxation studies, we
disconnect the thermalized array from the temperature-$T$ heat
bath (i. e., we remove the $\eta_i(t)$ and $\gamma_0$ terms
from the equations
of motion) and connect the ends of the chain (sites $1$ and $N$) via
a damping with rate $\gamma$ to a zero-temperature
reservoir.  This causes the total energy of the chain to decay
through these end points.  In all our simulations we set $\gamma=0.1$.
We present several sets of figures (all in dimensionless units)
to illustrate the relaxation behavior.
Typical energy relaxation curves for arrays averaged
over initial thermalized configurations are shown in
Fig.~\ref{figdecay}.  Associated with this evolution we define the
time-dependent spectra
\begin{equation}
S(\omega,t)\equiv 2\int_0^{\tau_{max}} d\tau~C(\tau,t) \cos \omega \tau
\end{equation}
where $\tau_{max}\equiv 2\pi/\omega_{min}$ and $\omega_{min}$ is chosen
for a desired frequency resolution; 
the choice $\omega_{min}=0.0982$, corresponding to $\tau_{max}=64$,
turns out to be numerically convenient. 
The time-dependent correlation function is actually an average over
the time interval $t-t_0$ to $t$, where we have
chosen $t_0 = 100$ (short enough for the correlation function
not to change appreciably but long enough for statistical
purposes) and is defined as follows:
\begin{equation}
C(\tau,t) =
\frac{1}{(N-1)}\sum_{i=2}^{N} \frac{1}{\Delta t}
\int_0^{\Delta t} d\tau ' ~\left< \Delta_i(t-\tau ') \Delta_i(t-\tau ' -
\tau) \right>,
\end{equation}
where $\Delta t \equiv t_0-\tau_{max}$. 
In Fig.~\ref{figsw} we display the evolution of the spectra for each
of the chains.  The spectral rendition is revealing because it clearly
indicates that the decay mechanisms for harmonic and each of the
anharmonic arrays are entirely different. 

%figure 3

\begin{figure}[htb]
\begin{center}
\epsfxsize = 4.in
\epsffile{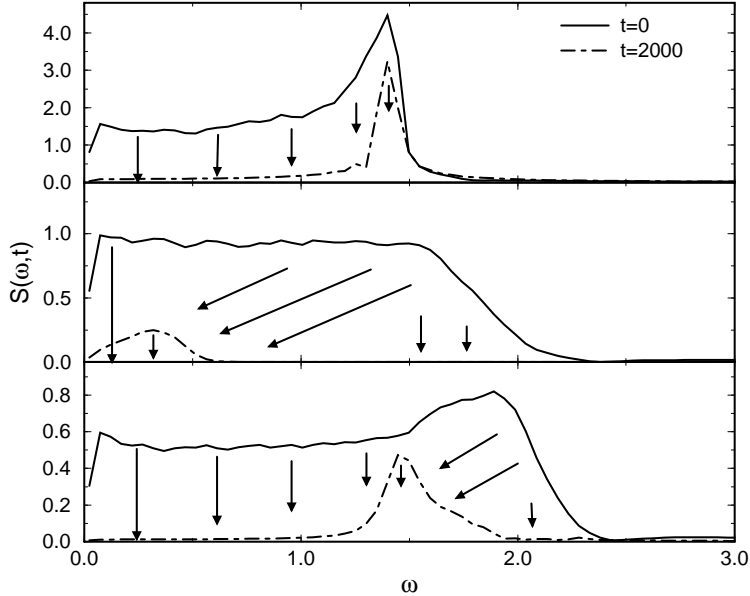}
\end{center}
\caption{Schematic representation of the spectral relaxation
channels.  The spectra of
Fig.~\ref{figsw} for times $t=0$ and $t=2000$ are shown again here, and the
arrows depict the pathways of different spectral components.  Downward
arrows indicate absorption by the cold reservoir, while angled arrows
denote degradation from one spectral region to another.  The
relative lengths of the arrows represent the associated rates.}
\label{figswesq}
\end{figure}

\begin{figure}[htb]
\begin{center}
\epsfxsize = 2.in
\epsffile{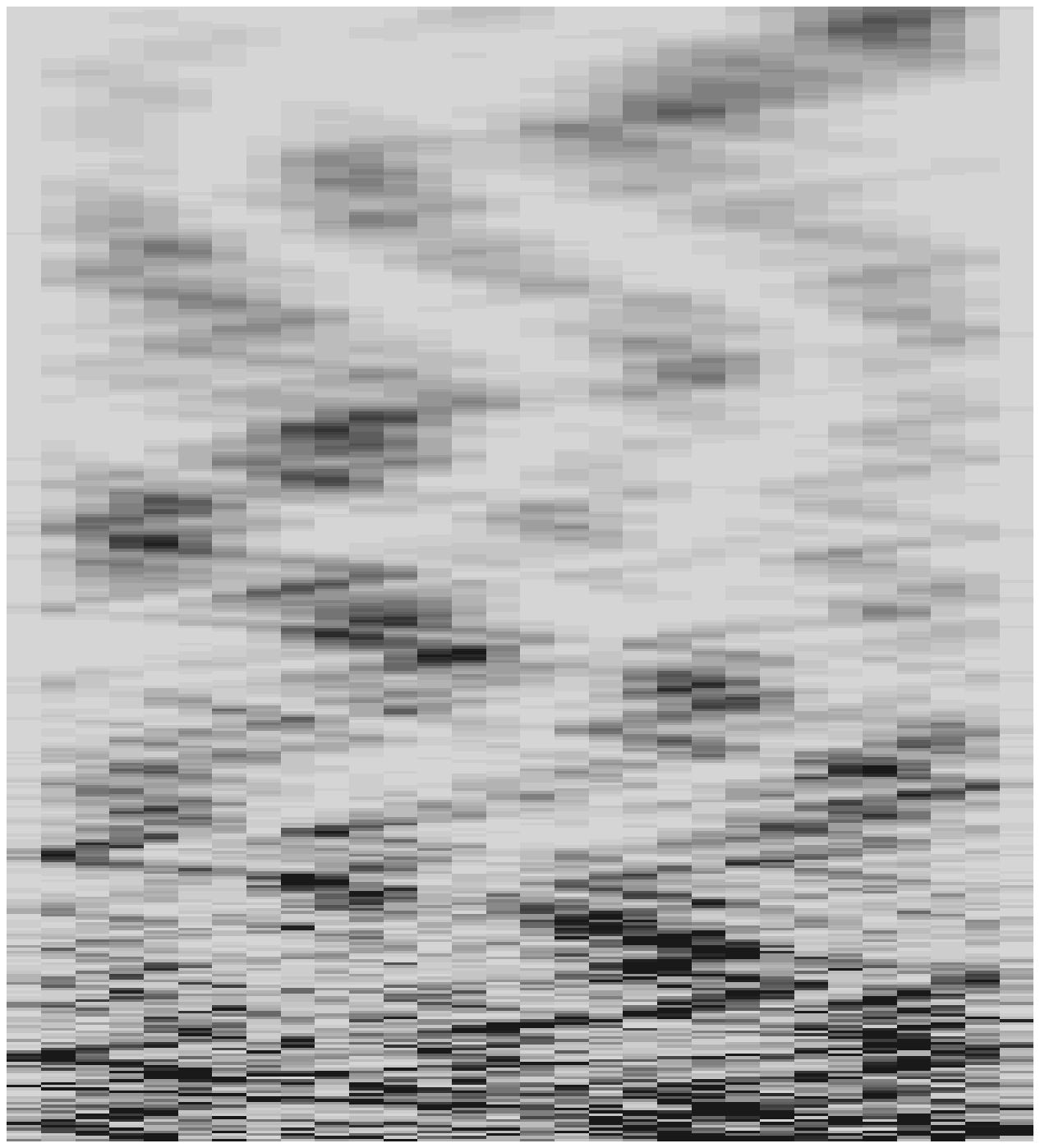}
\epsfxsize = 2.in
\epsffile{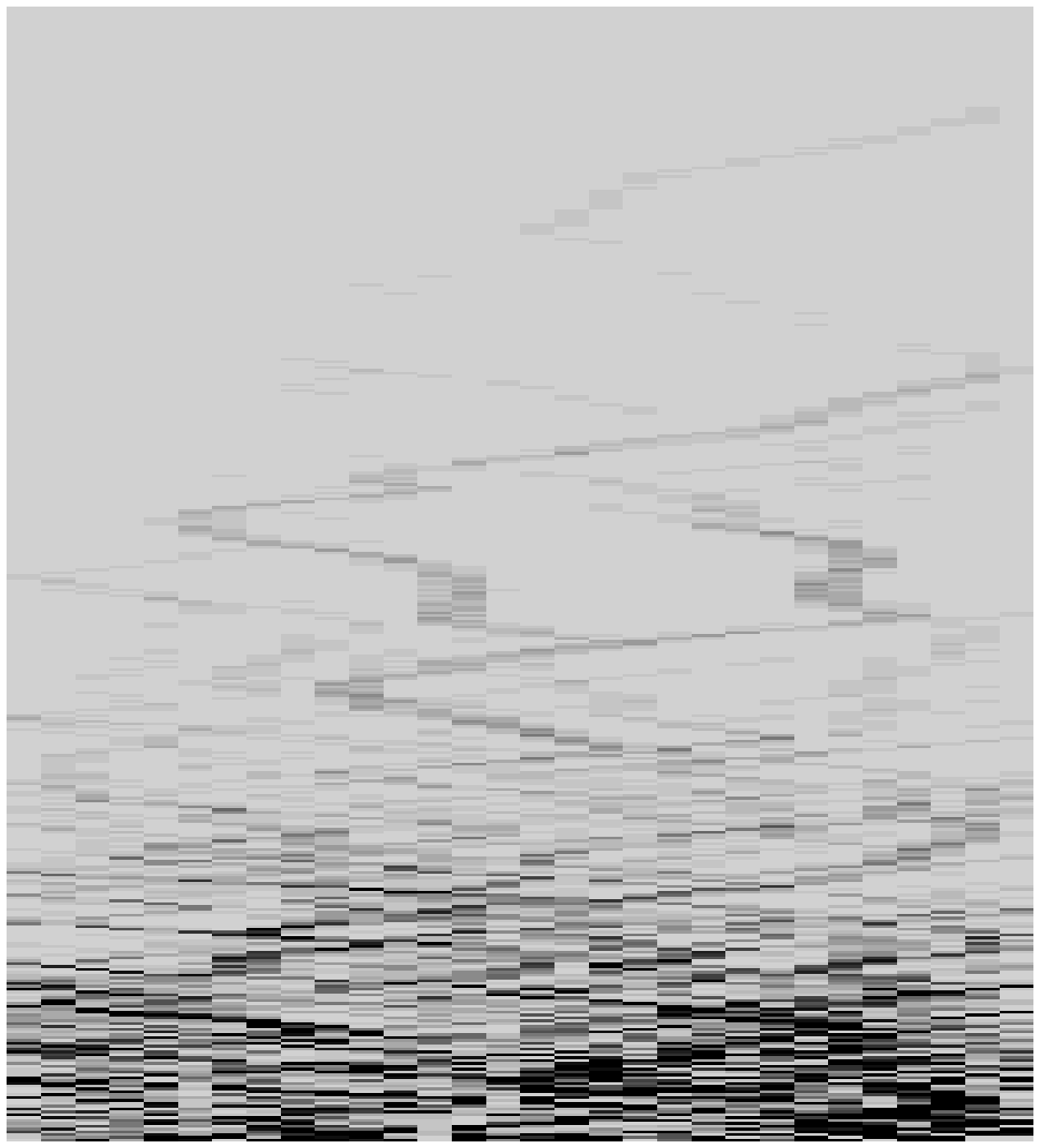}
\epsfxsize = 2.in
\epsffile{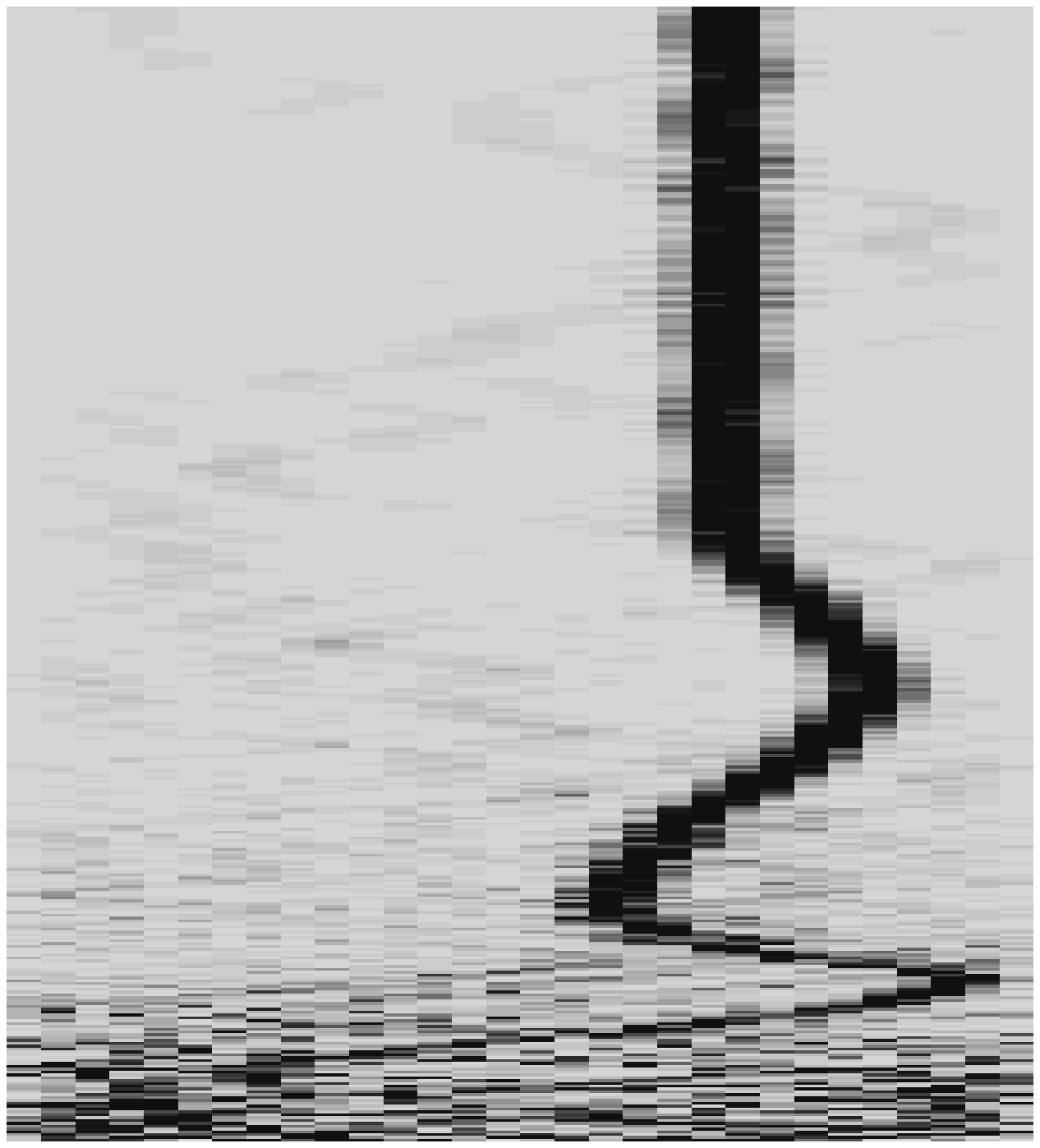}
\end{center}
\caption
{Energy landscapes of 30-site arrays initially thermalized at $T=0.5$.
Time advances along the $y$-axis until $t=1000$. 
A gray scale is used to represent the local energy,
with darker shading corresponding to more energetic regions. First panel:
harmonic
chain, $k=0.5$ and $k'=0$.  Second panel: purely anharmonic chain, $k=0$
and $k'=0.5$.
Third panel: mixed chain, $k=k'=0.5$.}
\label{figevol}
\end{figure}

Several features of the energy decay curves are noteworthy.
The energy in the harmonic array (solid curve in Fig.~\ref{figdecay})
is calculated by Piazza et al.~\cite{piazza} to be given by
\begin{equation}
\frac{E(t)}{E(0)}=e^{-t/\tau_0}I_0(t/\tau_0),
\end{equation}
where $I_0$ is the
modified zero-order Bessel function.  At first the decay is exponential
with time constant $\tau_0=N/2\gamma$ ($\tau_0=250$ in the figure),
and then at times larger than this time the decay changes to
$(t/\tau_0)^{-1/2}$.  The exponential decay time is that
associated with the lowest frequency phonon modes since they have
the shortest decay times.  The power law relaxation arises from a
cascade of different decay times of the different phonon
modes.  Eventually the decay becomes exponential again when
only the highest frequency modes survive, but the energy in our chain is
too low at that point to be picked up within our precision. 
A well-known but important point needs to be made here so that the
contrasting behavior of anharmonic chains can be clarified later:
the phonons in the harmonic chain are of course {\em independent}
of one another, and each has to be absorbed by the cold reservoir
separately. As noted above, each is absorbed on a different time scale.
In the first
panel of Fig.~\ref{figsw} we show the spectrum of the harmonic
chain at different times during the relaxation process, starting with the
initial thermalized spectrum (in this figure $S(\omega,0)\equiv S(\omega)$
of Fig.~\ref{figswt0}).
The evolution confirms that {\em low} frequencies decay more
rapidly in the harmonic chain -- the spectrum is
absorbed by the cold reservoir from the bottom up, and by time $t=2000$ 
only the longer-lived band-edge modes remain in the system.
The spectral decrease occurs ``vertically", that is, each spectral
component decays directly into the reservoir; this is shown schematically
in Fig.~\ref{figswesq}, where the downward arrows represent absorption by
the reservoir and their relative length schematizes the absorption rate.
The first panel of Fig.~\ref{figevol} shows the time evolution of the
local spatial energy landscape using a gray scale to contrast
higher-energy (darker) from lower-energy (lighter) regions.

%figure 4

The purely hard array relaxes more rapidly than the harmonic and
the energy decays essentially
exponentially, indicating a single predominant decay channel.
Note that being a purely
quartic chain, there are no phonons in this system. 
We also observe in the second panel of Fig.~\ref{figsw} 
that higher frequency excitations relax first.  At time $t=2000$ only 
low frequency excitations remain in the chain.  This behavior is exactly
opposite to that of the harmonic chain.
We find that
the dominant relaxation mechanism is for the high frequency portions of the
spectrum to degrade into lower frequency excitations, as schematically
indicated by the sloped arrows in  Fig.~\ref{figswesq}. Such a degradation
is possible here since individual frequencies are not associated with
normal modes in the anharmonic system.  In turn, 
these lower frequency excitations decay into the reservoir.  
Specifically, the high frequency components of the spectrum
are mainly due to mobile
localized modes that degrade into lower energy excitations
as they move and collide with one another, and this degradation
occurs with a relatively short time
constant that is shorter for localized modes that have a higher velocity.
The lower frequency excitations are in turn absorbed
into the cold reservoir but continue to be replenished through the
degradation process. We conjecture that the absorption of
the lowest frequency components defines the observed exponential decay
constant of the total chain energy.
However, and importantly, among the low frequency excitations are some
that persist for a very long time, certainly beyond the times of
our simulations. 
These, which are the only remaining spectral components at time $t=2000$,
are ``labeled" by short downward arrows in the relaxation
schematic and include rather stable breather
and/or soliton modes that move very slowly and are localized away from the
boundaries.   At the same time, some portion of the high frequency spectrum
is also directly absorbed by the cold reservoir (indicated by the
short downward arrow in the high frequency region of
Fig.~\ref{figswesq}).  For example, 
when a highly mobile localized excitation reaches a boundary it may
be absorbed directly by the reservoir (or it may be reflected and return
into the chain with some energy loss).  
The role of high-frequency mobile modes and of low-frequency 
slowly moving or stationary modes
in this picture will be tested in more detail in the next section,
where we explicitly inject a high-frequency localized mode into the array
and observe the relaxation 
dynamics. We do note here that our picture is consistent with
known facts about localized states.  In particular, it is known that 
higher-frequency and/or higher amplitude localized modes can
move at higher velocities~\cite{bourbonnais,cretegny,kosevich}.
It is also known that while in motion such modes lose energy through
collisions with other excitations. 
Also, Fig.~\ref{figdecay} shows a faster decay at higher temperatures,
which is consistent with our observations elsewhere that the speed
of an injected pulse (and therefore, we conjecture, the speed of a 
moving localized mode) in these arrays increases with
temperature~\cite{wepulse}.  The energy
landscape associated with these descriptions is shown
in the second panel of Fig.~\ref{figevol}.  

%figure 5
\begin{figure}[htb]
\begin{center}
\epsfxsize = 4.in
\epsffile{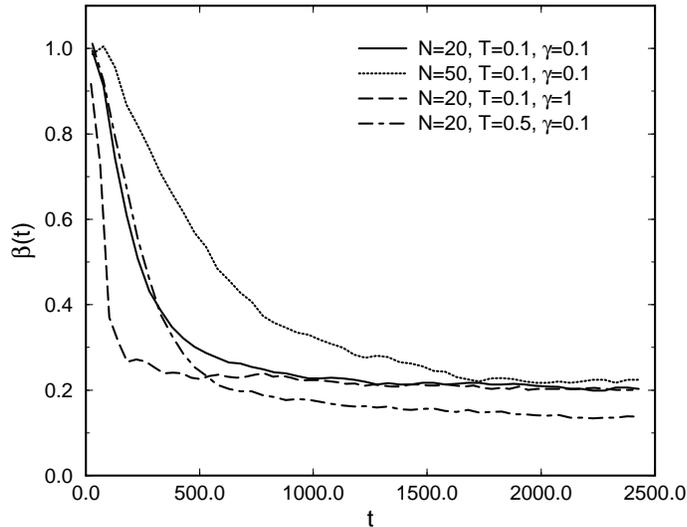}
\end{center}
\caption{Plot of $\beta(t)$ as a function of time for various mixed arrays. 
A flat line below $\beta(t)=1$ indicates stretched exponential behavior.}
\label{figmixto}
\end{figure}

Almost all FPU analyses in the literature involve arrays that include both
harmonic and anharmonic contributions, but the distinct role of each
has not been clarified.  Our relaxation results show an interesting
sequence of relaxation behaviors.  At early times the mixed array
relaxes more rapidly than the harmonic, because there are
low-frequency excitations close to harmonic phonons
(but note that phonons are no longer exact normal modes)
{\em and} high-frequency excitations in the system. Energy relaxation
and decay thus involves {\em both} of the mechanisms discussed above.
Again, because initially the high-frequency modes move more rapidly
at higher temperature, the early time decay is faster at higher
temperatures.  That both low and high frequency modes relax rapidly 
is clearly seen in the third panel of
Fig.~\ref{figsw}, which quickly loses both low (as in the first panel)
and high (as in the second panel) frequency
portions of the spectrum.  In the energy decay curve there is then a
crossing after which the mixed chain relaxes much more {\em slowly} than
the harmonic and the purely anharmonic.  This occurs when the low frequency
modes (phonons) have essentially all decayed, and only certain
high-frequency spectral components remain, as clearly seen in the spectrum.
We conjecture that these persistent modes are
localized {\em quasi-stationary} breathers that decay extremely slowly,
and again pursue this notion in the
next section.  The associated decay schematic is illustrated
in Fig.~\ref{figswesq}.  Note that with increasing temperatures
the stationary breathers are more energetic, leading to a slower
decay of the total chain energy at long times.  Indeed, we find that the
very slow decay at long times is a stretched exponential, as shown in
Fig.~\ref{figmixto}.  Here we plot
\begin{equation}
\beta(t) = \frac{d}{d\ln t} \ln \left[ -\ln
\left(\frac{E(t)}{E(0)}\right)\right].
\end{equation}
If $E(t)/E(0)$ is of stretched exponential form $\exp[-(t/\tau)^\sigma]$,
then $\beta(t)=\sigma$.  Figure~\ref{figmixto} shows the stretched
exponential behavior for various arrays and clearly points to a
$T$-dependent but $N-$ and $\gamma$-independent exponent $\sigma$.  The
decrease of $\sigma$ with increasing initial temperature is explained by
the greater stability of more energetic breathers.  The $N$ and $\gamma$
independence is explained by the fact that the rate-limiting step in the
slow relaxation is the leakage by the breather.  The low-energy output of
this leakage is quickly absorbed by the cold reservoir~\cite{tsi2}.

These results immediately raise a question that in hindsight pervades a
number of results throughout the literature: if both the purely quartic
chains and also the mixed chains support high-frequency localized solutions, 
why do these modes relax so rapidly in the former
but seem to persist for very long times in the latter?  The answer
lies in the crucial role of the
quadratic terms of the potential and the consequent behavior of the
low-frequency portion of the spectrum.  When localized solutions (breathers)
are sufficiently strongly perturbed, they 
respond by moving~\cite{bourbonnais,cretegny,kosevich} and hence to
the degradation process described earlier.  In
a mixed chain, the low-frequency excitations 
(phonons) that can easily perturb breathers decay quickly,
and at later times only
the essentially unperturbed high-frequency breathers remain. 
These almost-stationary
solutions of the system hardly move.  Since localized breather modes tend to
lose their energy only while they move, these quasi-stationary breathers 
can persist for a very long time.  In the purely anharmonic chain, on
the other hand, there are no phonons and the low-frequency
excitations include slowly-moving quasistationary anharmonic modes
that persist for a long time and that continue to perturb
the high frequency localized modes.  These energetic localized
modes thus continue to move and collide with the low-frequency modes,
resulting in degradation into lower energy excitations.

The third panel in
Fig.~\ref{figevol} shows the spontaneous appearance, slowing motion, and
eventual stoppage of a breather in the mixed chain.  It is this breather,
absent from the purely hard array,
that mainly leads to the persistent high-frequency spectral
contributions and to the extremely slow relaxation of the mixed
chain energy at long times.

%figure 6

\section{Relaxation of Thermalized Array With an Injected Localized
Excitation}
\label{injected}

In the previous section we portrayed a relaxation dynamic for anharmonic
FPU arrays that involves a very specific view of the roles of
high-frequency localized modes, low-frequency anharmonic modes, and
phonons.  In order to further test these ideas, 
in this section we start again with the thermalized chain,
but now we inject a high-amplitude localized excitation
at time $t=0$ in the center of the chain. 
Specifically, we inject an odd-parity excitation (amplitude $A$ at site
$N/2$ and $-A/2$ at each immediately adjacent site).  These
displacements lead to an exact breather solution for the interaction
potential $V(x_i-x_{i-1})
=(x_i-x_{i-1})^n$ as $n\to\infty$ (as does the even-parity breather of
amplitude $A$ at one site and $-A$ at an immediately adjacent site),
and are quite close to exact for
the quartic FPU potential~\cite{flach,MA}.  The fate of
the excitation as the entire system relaxes clarifies the
roles of the different spectral components in the relaxation process.
The excitation amplitude is
sufficiently large ($A=2$) to insure clear presence above the thermal
background. 

In the previous section we introduced the notion of localized modes
as an important component in the thermal relaxation process of the
FPU systems.  We explicitly differentiated between {\em mobile} and {\em
stationary} localized modes, and noted that energy loss
occurs when a localized mode moves and collides with other localized modes.
We also stated that this energy loss
occurs through degradation into lower-frequency modes.  In order to focus
on this mechanism without additional interference from the ends of the chain
other than the normal low-frequency decay processes discussed earlier,
in this section we use relatively long chains, $N=300$.  This is
sufficiently long that we never see a high-energy localized
mode reaching a boundary site before it has degraded or stopped moving.

\begin{figure}[htb]
\begin{center}
\epsfxsize = 4.in
\epsffile{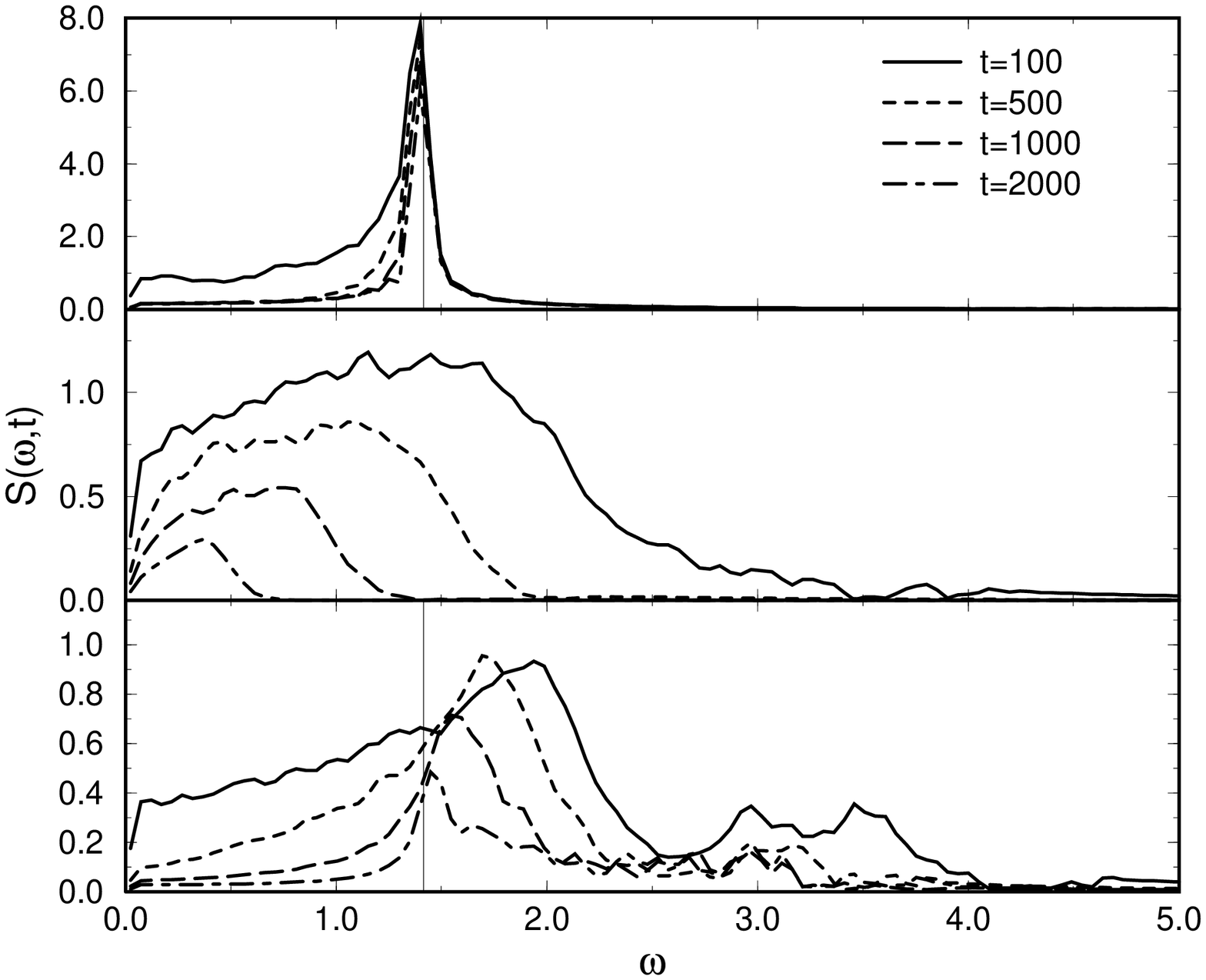}
\end{center}
\caption{Time evolution of spectra for various relaxing
arrays initially at $T=0.1$ with a high-amplitude
localized mode injected at $t=0$.  
First panel: harmonic chain with $k=0.5$.  Second panel: purely
anharmonic chain with $k'=0.5$.  Third panel: mixed chain with $k=k'=0.5$.
The thin vertical line again indicates the frequency
$\omega=\sqrt{4k}=\sqrt{2}$.  The $t=0$ spectra are not shown.}
\label{figbre}
\end{figure}

The motion of the injected excitation during the relaxation process
is followed in two ways. As before, we obtain
the spectrum of the chains at various times (see
Fig.~\ref{figbre}). We also calculate the mean
squared displacement
\begin{equation}
\left< x^2(t)\right> \equiv \left< (i_{max}(t) - \frac{N}{2})^2\right>
\end{equation}
as a measure of the position of the excitation
(its dispersion in the anharmonic chains is very
small~\cite{wepulse}).  Here $N/2$ is the initial point of
highest energy in the chain and $i_{max}(t)$ is the point of maximum energy
at time $t$.

In a harmonic chain the progression is as expected for a harmonic system.
The initial localized excitation spreads quickly over the entire
array and thus loses its localized character. The associated Fourier
decomposition into phonon modes dictates the relaxation behavior
seen in Fig.~\ref{figbre}, which is much like that seen in
Fig.~\ref{figsw}  except that the high-frequency (longer-lived) phonon
modes are now more populated.  

%figure 7

\begin{figure}[htb]
\begin{center}
\epsfxsize = 4.in
\epsffile{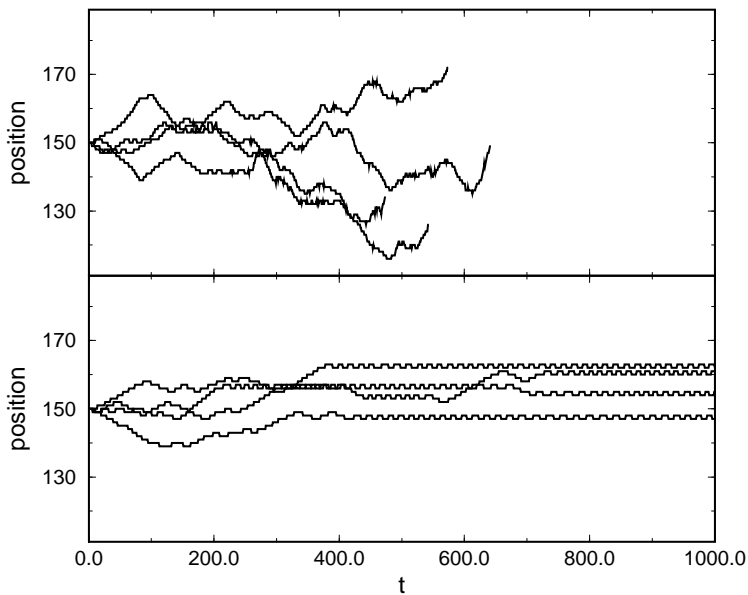}
\end{center}
\caption{Typical injected breather trajectories. Upper panel: purely
anharmonic array.  Lower panel: mixed array.}
\label{figtraj}
\end{figure}

In the purely anharmonic chain, shown in the second panel of
Fig.~\ref{figbre}, the initial excitation introduces frequency
components in a fairly broad spectral range (including frequencies
well above the corresponding thermal range). Part of
this spectral contribution is associated with an excitation that
remains spatially localized (the rest appears because the injected
excitation is not an exact mode of the thermalized chain;
whereas the localized mode appears with each realization, the other
spectral contributions vary somewhat in detail from one realization
to another).  Consistent with our description
in Sec.~\ref{thermal}, the high
frequency components again relax quickly, indicating an
energy degradation
of the high-frequency excitations into lower frequency modes. 
The detailed trajectory of the initial localized excitation
is quite interesting, and
a few particular realizations are shown in the upper panel
of Fig.~\ref{figtraj}.  
After a short time the excitation begins to move in one direction or the
other with equal probability
(an initially even-parity excitation behaves very similarly but takes a
longer time to begin to move because even-parity breathers are more
stable in FPU chains).  The motion continues for a period of random
duration.  Then the excitation stops moving for a random period,
until it moves again in either direction for a random period of time. 
While stationary, we have observed that (whatever its initial
configuration, even or odd) the excitation has
even parity, but when it moves it alternates between even
and odd parity.  Furthermore, the excitation only loses energy while in
motion.  A detailed analysis reveals that the stationary excitation is
perturbed again and again by slow low-frequency localized
excitations that collide 
with it and repeatedly set it in motion.  These low frequency modes are
precisely those that, as described in the previous section, persist
for a long time in the absence of a harmonic component. This sequence of
events serves to confirm our thermalization analysis of Sec.~\ref{thermal}. 

%figure 8

%figure 9
\begin{figure}[htb]
\begin{center}
\epsfxsize = 4.in
\epsffile{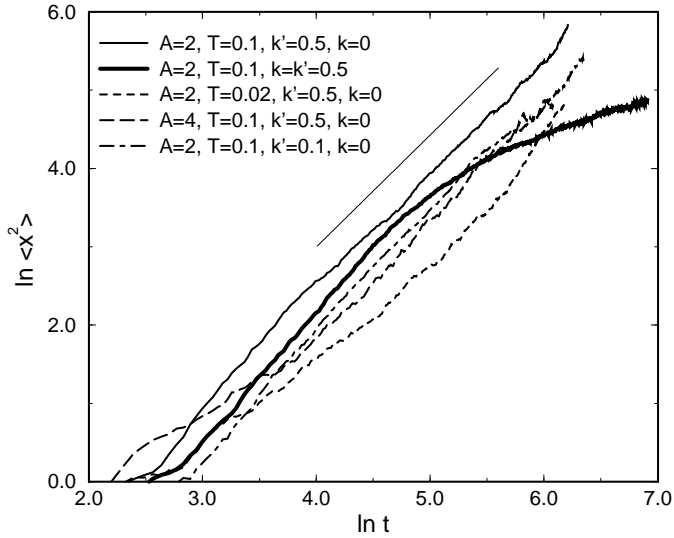}
\end{center}
\caption{Mean squared displacement of a localized mode in various FPU
arrays at various temperatures. The short straight line is a guide
to the eye and has a slope of 3/2.}
\label{figxxb}
\end{figure}

Mean squared displacement results for a variety of parameter combinations
are shown in
Fig.~\ref{figxxb}.  The mean squared displacement is seen to follow
the superdiffusive law 
$\left< x^2(t)\right> \sim t^{\alpha}$ with $\alpha=3/2$ over the entire
lifetime of the excitation.  This particular exponent is recovered for the
purely quartic chain under all conditions that we have tested, that is,
independently of force constant, excitation amplitude, and temperature. 
Variations in parameters affect the breather velocity, which in turn
modifies the coefficient of $t^{3/2}$, but not the power (a higher
temperature, a stronger force constant, and a higher excitation amplitude
all lead to higher velocities).
Indeed, it does not even matter {\em when} in the course of
the relaxation process the
localized excitation is introduced: its mean squared displacement follows
the above behavior until the excitation is extinguished, again confirming
that this behavior is mainly caused by the persistent low-frequency
excitations. 

%figure 10
\begin{figure}[htb]
\begin{center}
\epsfxsize = 4.in
\epsffile{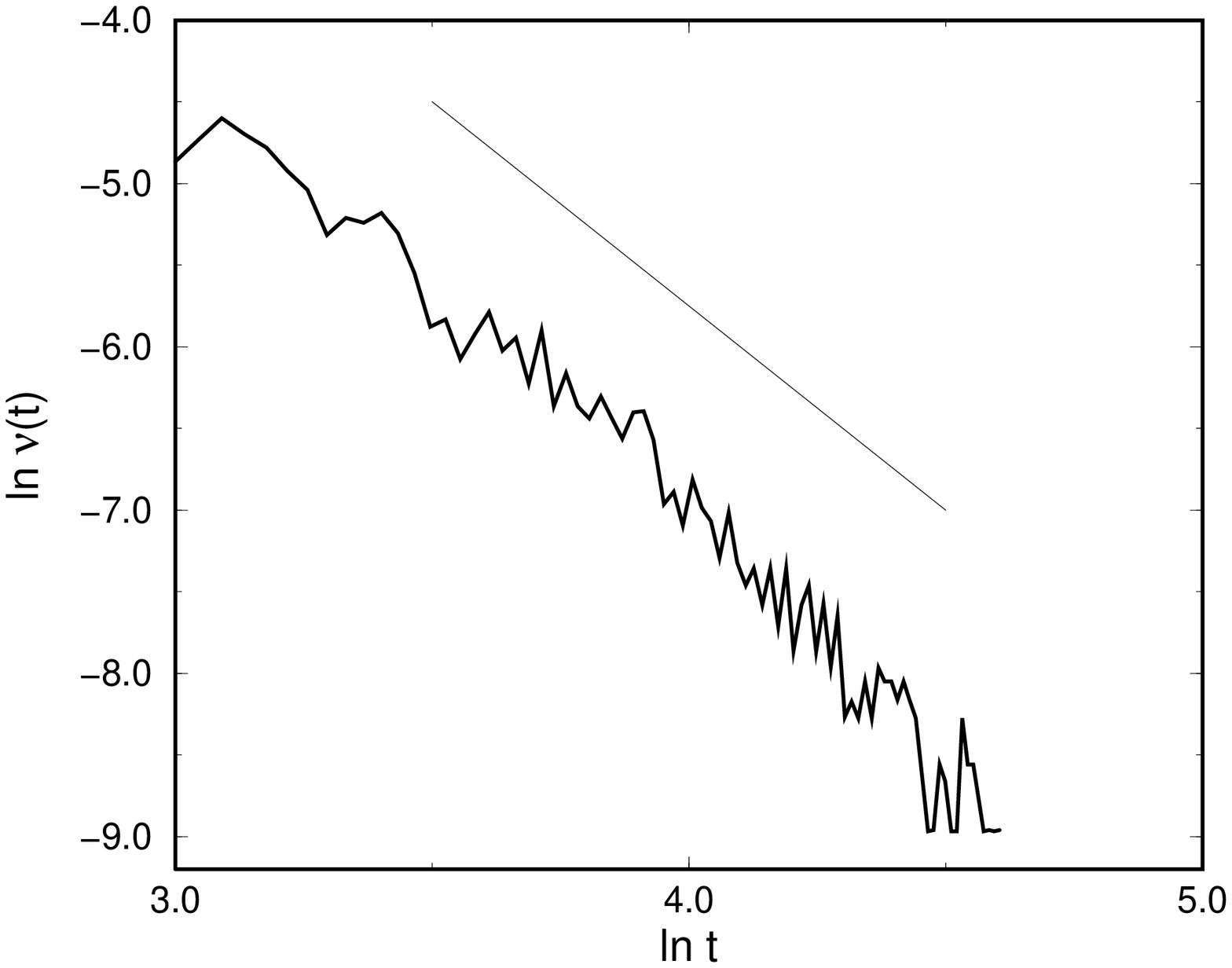}
\end{center}
\caption{Distribution of collision times of the high-frequency localized
mode in the purely anharmonic FPU chain for times much greater than the
typical oscillation period of the breather.  The straight line has slope
$-5/2$.} 
\label{figreb}
\end{figure}

A model that leads to the observed power law and contains the main features
of the excitation collision picture was recently developed in a different
context~\cite{barkai}.  It describes a light particle that
moves with a constant speed $\pm v$ among point scatterers arranged
randomly on a line.  The intervals between scattering points, $\xi_n$, are
independent identically distributed random variables described by a
probability density function $\mu(\xi)$.  If the density decays as
$\mu(\xi)\sim \xi^{-(1+\gamma)}$ with $1<\gamma<2$ when $\xi$ is large,
then the mean squared displacement of the light particle goes as $\left<
x^2\right> \sim t^\alpha = t^{3-\gamma}$.  In particular, when
$\gamma=3/2$ then $\alpha=3/2$ as in our results.  This model suitably
modified may describe our system.  Our scatterers (low-energy breathers)
do move, but all that is required to obtain the observed power law is that 
the times at which they collide with the injected excitation
be distributed according to $\nu(t)\sim t^{-5/2}$ (in the work of Barkai et
al.~\cite{barkai}, distance intervals and time intervals are
interchangeable because the scatterers are stationary).  The
quiescent periods
simply modify the coefficient of this and of the resulting mean squared
displacement distribution, but the power $\alpha$ is determined by the
collision time distribution exponent.  Numerical results for the
distribution of times between breather collision events are shown in
Fig.~\ref{figreb}.  The data is noisy (and could of course be made smoother
with more realizations), but the confirming trend is clear.
This is of course simply
phenomenology, since we have no explicit dynamical model to obtain this
distribution.  

In the mixed chain, whose spectra are shown in the third panel of 
Fig.~\ref{figbre}, the initial excitation again leads to the appearance
and persistence of high frequency spectral components.
Typical trajectories of the highest energy modes (which here, too, remain
localized) are shown in  Fig.~\ref{figtraj}.  The
difference between the purely anharmonic and the mixed typical trajectories
are evident: whereas the excitations in the former continue to move until
extinguished, the excitations in the mixed chain slow down and eventually
stop altogether when all perturbing excitations have been swept out
of the system.  Once there are essentially no other excitations to
collide with a stationary breather, it remains in the system in
spite of the dissipative bath acting on its ends.   This is completely
consistent with the landscape shown in Fig.~\ref{figbre}.
The mean squared displacement in
Fig.~\ref{figxxb} clearly reveals this behavior as well: the exponent
begins at $3/2$ but eventually bends towards $\alpha=0$ when
the breather stops moving.

\section{Conclusions}
\label{summary}

In this paper we have studied energy relaxation in one-dimensional
nonlinear arrays with quartic interparticle interactions (Fermi-Pasta-Ulam
or FPU arrays). In one scenario,
we have thermalized the arrays to a temperature $T$ and then observed the
relaxation of the arrays when the boundaries are connected to a
zero-temperature reservoir through damping terms.  In another scenario, we
have introduced a high-energy localized excitation in the thermalized
array and have observed the relaxation process and the fate of this
excitation. This second scenario serves to confirm our description
of the dynamics of thermal relaxation.  Throughout we have
applied free-end boundary conditions.

Our most salient results concern the role of harmonic contributions to the 
FPU potential as distinct from the thermalization of an array.   In other
words, we emphasize that one can equilibrate an array at a given
temperature whether or not its interactions include quadratic
interactions, and we pose the following question: What exactly is
the role of these interactions?  

We have confirmed that thermal relaxation in a purely harmonic chain 
involves the sequential decay of independent phonon modes starting
with those of lowest
frequency and moving upward across the spectrum.  We have
also confirmed that the total energy of the array
decays exponentially for short times and as an inverse power law at longer
times, as calculated by Piazza et al.~\cite{piazza}.  When a localized
excitation is introduced in a purely harmonic array it quickly spreads and
loses any localized identity. 

In a purely anharmonic chain there are no phonons, and the anharmonic
excitations in general include localized modes.  High-frequency
spectral components include highly localized modes that may be stationary
but are easily set in essentially ballistic motion by sufficiently
strong scattering events.  As a result, their net motion is
superdiffusive.
When not in motion these modes can retain their energy for a long time,
but while in motion they lose energy
through collisions with other excitations
and eventually degrade into lower energy excitations.  The lowest
energy excitations decay into the cold reservoir, while other
low-energy excitations persist for a long time in the chain.  
The spectral relaxation proceeds mostly from the high frequency end of the
spectrum downward.  At long times the energy residue that remains
in the chain in the form of low-frequency localized excitations
that move slowly is quite persistent but very small.  
To confirm this description we have observed the dynamics
of an injected localized high-amplitude excitation.  We observe that it is 
perturbed by the thermal excitations, which sets the breather in motion. 
This motion alternates with quiescent periods, but resumes when
the excitation is again perturbed sufficiently strongly.
Since during its lifetime there is always a slowly moving thermal
background, the breather continues to resume motion until
it disappears into the relaxing thermal background.  The time-dependence of
the mean squared displacement of the breather is remarkably universal over
its entire lifetime, $\left< x^2\right> \sim t^{3/2}$, independently
of initial breather amplitude, temperature, and force constant.

In a mixed anharmonic array the relaxation process
involves phonon-like modes (with the lowest frequencies decaying first)
and also high-frequency anharmonic modes (with the highest frequencies
decaying first).  The
relaxation is at first rapid, but as the phonon decay ``sweeps" the system
clean of low-energy excitations, quasi-stationary high energy breathers
are no longer perturbed and remain essentially stationary; the
subsequent relaxation process is exceedingly slow. 
When a high-energy localized excitation is injected in
the mixed array,
at first it is perturbed by thermal excitations that induce motion.
However, as the thermal excitations are swept out of the system through the
harmonic channel, the breather stops moving and survives for a very long
time, thus confirming the relaxation picture.  Associated with
this description is a mean squared displacement that
at first goes as $\left< x^2\right> \sim t^{3/2}$ but
then becomes independent of time.

From a broader perspective, we have shown that vibrational energy
localization and persistence is aided by the presence of an efficient
mechanism to remove other background excitations that might perturb and/or
destroy localization.  In our specific model, localization is due to hard
anharmonic interactions and the removal mechanism involves a harmonic
phonon channel, but one can envision other localization and background
sweeping processes that lead to a similar outcome.  From a narrower
perspective, there are a number of questions that remain to be explored,
including relaxation in higher dimensions and the incorporation of more
realistic potentials.

\section*{Acknowledgments}
This work was supported in part by the
Engineering Research Program of the Office of Basic Energy Sciences at
the U. S. Department of Energy under Grant
No. DE-FG03-86ER13606. Partial support was provided by a grant from the
University of California Institute for M\'exico and the United States (UC
MEXUS) and the Consejo Nacional de Ciencia y Tecnolog\'{\i}a de
M\'exico
(CONACYT), and by IGPP under project Los Alamos/DOE 822AR.

%\end{multicols}

\end{document}